\documentclass[fleqn,10pt]{wlscirep}

\usepackage{amsmath,xcolor,amsthm,amssymb,bm}
\usepackage{siunitx}
\usepackage{graphicx}
\usepackage{xcolor}
\usepackage{comment}

\newcommand{\dd}[1]{\mathrm{d}#1\,}

\DeclareMathOperator{\sech}{sech}

\newcommand{\ph}{\varphi}

\newcommand{\DT}{\Delta(T)}

\newcommand{\en}{\epsilon}
\newcommand{\be}{\begin{equation}}
\newcommand{\ee}{\end{equation}}

\title{Coherent Josephson thermodynamic cycles}
\author[1, 2*]{Francesco Vischi}
\author[2]{Matteo Carrega}
\author[2]{Pauli Virtanen}
\author[2]{Elia Strambini}
\author[2]{Alessandro Braggio}
\author[2]{Francesco Giazotto}
\affil[1]{Dipartimento di Fisica, Universit\`{a} di Pisa, I-56127 Pisa, Italy}

\affil[2]{NEST, Istituto Nanoscienze-CNR  and Scuola Normale Superiore, I-56127 Pisa, Italy}
\affil[*]{francesco.vischi@df.unipi.it}

\begin{abstract}
A superconductor/normal metal/superconductor Josephson junction is a coherent electron system where the thermodynamic entropy depends on temperature and phase difference across the weak-link.
Here, exploiting the phase-temperature thermodynamic diagram of a thermally isolated system, we argue that a cooling effect can be achieved when the phase drop across the junction is brought from 0 to $\pi$ in a {\it iso-entropic} process. We show that {\it iso-entropic} cooling can be enhanced with proper choice of geometrical and electrical parameters of the junction, i.e. by increasing the ratio between supercurrent and total junction volume. We present extensive numerical calculations using quasi-classical Green function methods for a short junction and we compare them with analytical results. Interestingly, we demonstrate that phase-coherent thermodynamic cycles can be implemented by combining {\it  iso-entropic} and {\it iso-phasic} processes acting on the weak-link, thereby engineering the coherent version of thermal machines such as engines and cooling systems. We therefore evaluate their performances and the minimum temperature achievable in a cooling cycle. 
\end{abstract}
\begin{document}
\flushbottom
\maketitle

\thispagestyle{empty}

\section{Introduction}
Recently, a great interest focussed on the study of mesoscopic hybrid devices with superconducting elements, both from a fundamental and technological point of view. Important applications of these systems can be found in the development of new single photon detectors\cite{giazotto2008,solinas2017,govenius2016}, fast and precise thermometers\cite{gasparinetti2015,Giazotto2016,zgirski2017,wang2018}, sensitive magnetometers\cite{ronzani2014,jabdaraghi2018, vischi2017, bours2018}, low-temperature phase-coherent caloritronics\cite{fornieri2017,martinezperez2014,timossi2018,fornieri2017_2,Guarcello2018,Guarcello2018b} and cryogenic machines\cite{marchegiani2017,muhonen2012,giazotto2006, savin2004,giazotto2014,giazotto2004,roddaro2011}. To this end a deep understanding of thermodynamic aspects of mesoscopic hybrid structures is timely and of upmost importance.

Moreover, low-temperature applications, micro-refrigerators operating at sub-Kelvin range are keypoint both for quantum technology and for astrophysics purposes\cite{astrophysics2010}. While it is rather easy to cool down  a system until few Kelvin, investigation of system at lower temperature is not a trivial task. Today the most common approach to achieve temperature of the order of tens of mK uses adiabatic demagnetization or dilution fridges, which however have  several limiting factors such as their bulky nature, their multi-stage operation and their very high-cost. Alternative cooling schemes are thus required to bypass these bottlenecks. Several systems at the microscale have been proposed to implement thermal machines, i.e. by using quantum dots\cite{prance2009}, single ions\cite{rossnagel2016}, microelectromechanical systems\cite{whalen2003}, piezoelectric elements\cite{steeneken2013}, or by exploiting adiabatic magnetization of superconducting structures\cite{dolcini2009,yaqub1960}.

State-of-the-art cooling microsystems have been realized in hybrid structures, using a normal metal/insulator/superconductor (NIS) junction\cite{nahum1994,giazotto2006}. Once voltage biased, they can be used as cooler or heater, resulting in substantial electronic cooling power if operated in the optimal regime with a voltage very close to the superconducting energy gap. The operating principle of these systems relies on the presence of the superconducting gap in order to remove the most energetic quasiparticles from the system\cite{lowell2013,oneil2012,vasenko2010,giazotto2006}. The performance of these systems have been further improved by connecting two junctions in series to realize  a SINIS structure\cite{leivo1996,nguyen2016,pekola2004}.

In this work, we investigate the possibility to achieve  cooling with a SNS Josephson weak-link. Differently from SINIS structures, here the mechanism relies on proximity effect\cite{likharev1979, pannetier2000,courtois1999}, which  takes place when a normal metal is in good electric contact with a superconductor. This allows to develop a phase-tunable minigap in the quasi-particle Density of States (DoS) of the normal region\cite{lesueur2008,giazotto2009,belzig1999,hammer2007}, when the weak-link is shorter or comparable to the superconducting coherence length $\xi$. This results in a non-trivial phase dependence of thermodynamic quantities, such as the entropy or the specific heat of the SNS junction. 
Here we exploit this phase dependence in an isolated system to envision a quantum iso-entropic process. Within this process the electronic temperature in the junction will be driven by the phase difference in order to conserve the entropy. 
In particular, we will show that it is possible to realize a cooling process based on this principle, similar to the adiabatic magnetic cooling\cite{dolcini2009,yaqub1960}.

The thermal isolation necessary for the iso-entropic process requires low operating temperatures of our cooler in order to decouple the electronic system from the phonon bath\cite{timofeev2009,muhonen2012,giazotto2006}, as already demonstrated in many advanced nanotechnologies \cite{koppinen2009,clark2005,nguyen2016}.
  
 Our phase-tunable iso-entropic process defines a new quantum thermodynamic building-block that can implement novel thermodynamic cycles. As an example we will combine two iso-entropic with two iso-phasic processes to realise a {\it thermodynamic   Josephson  cycle}. We therefore investigate in details the performances of realistic thermal machines based on this quantum coherent cycle both as engine and cooler.
 
The paper is organised as follows. In Section \ref{sec:model} we introduce the model and basic definitions of thermodynamic quantities. 
In Section \ref{sec:iso-entropic} we show the possibility of performing an iso-entropic thermodynamic process in a SNS Josephson junction. 
Iso-entropic and iso-phasic processes are combined to form a thermodynamic Josephson cycle, as we will discuss in Secction \ref{sec:cycles}. 
Finally, Section \ref{sec:exp} contains some comments on possible experimental realizations, while Section \ref{sec:conclusion} summarize our main findings.

\section{Model and thermodynamic quantities}
\label{sec:model}
\begin{figure}
  \includegraphics[width=0.99\textwidth]{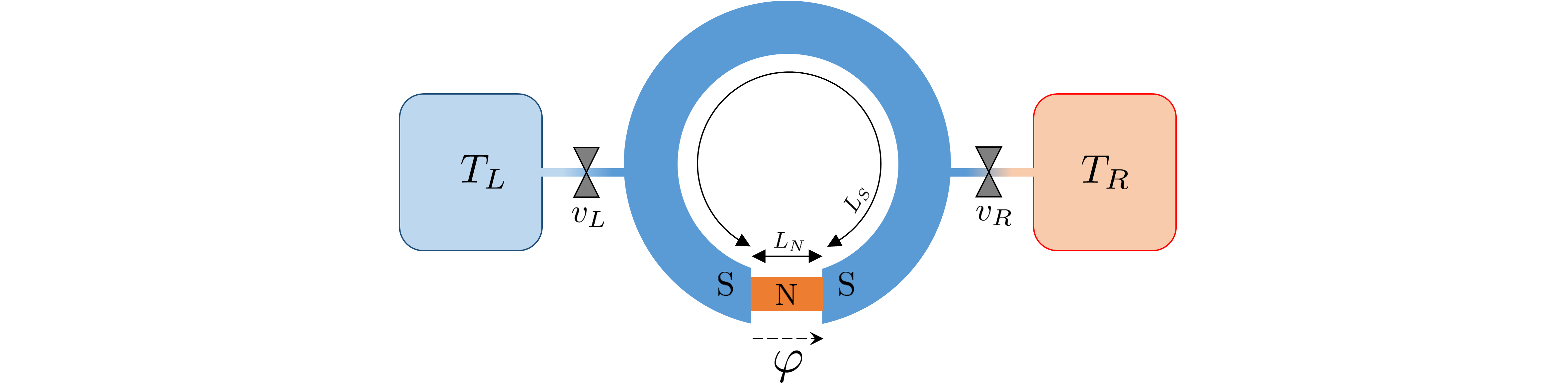}
  \caption{\textbf{ Thermodynamic SNS system.}  A superconductor/normal metal/superconductor Josephson weak-link is phase-biased at phase $\ph$ by a superconducting ring pierced by a magnetic flux. The electronic system can be connected via two thermal valves $v_L$ and $v_R$ to two external reservoirs residing at temperature $T_L$ and $T_R$, respectively. $L_N$ and $L_S$ are respectively the junction and ring lengths}
  \label{fig:sketch}
\end{figure}
\begin{figure}
  \includegraphics[width=0.99\textwidth]{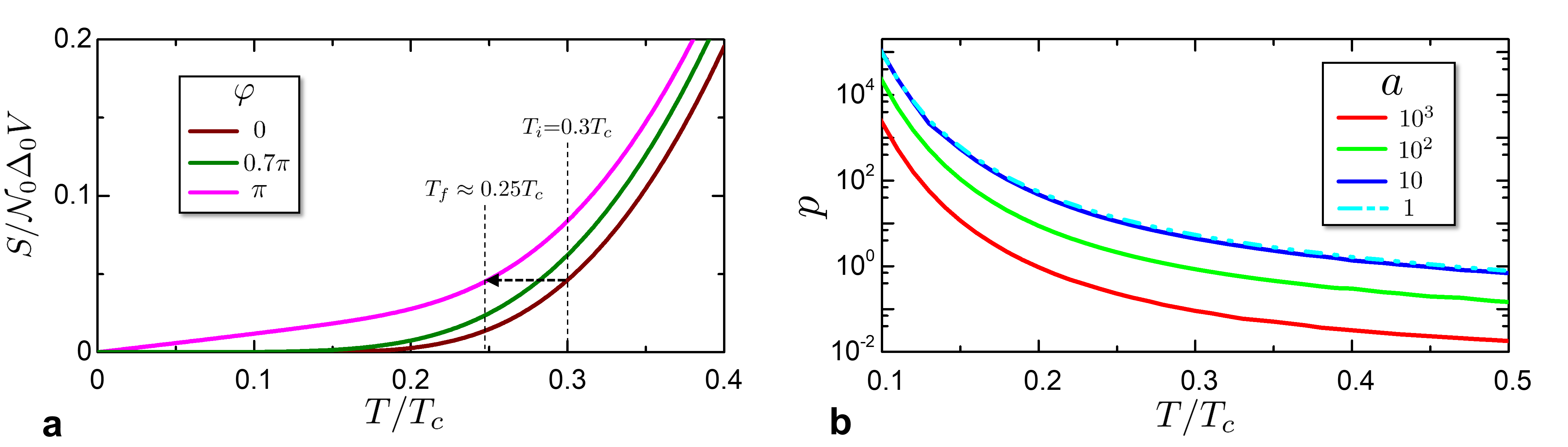}
  \caption{\textbf{ Temperature dependence of the entropy of the SNS system.} (a) Entropy $S(T,\varphi)$ as a function of temperature $T$ (in units of the critical temperature $T_c$) for three representative values of phase. The arrow shows an iso-entropic process from $\varphi=0$ to $\varphi=\pi$ and the corresponding electronic temperature variation from $T_i$ to $T_f$. In this plot the proximity parameter $a=(\sigma_S A_s)/(\sigma_N A_N)=100$. (b) Relative entropy variation $p$ as a function of $T/T_c$ for different values of $a$.
}
  \label{fig:two}
\end{figure}
We consider a superconductor/normal metal/superconductor (SNS) Josephson weak-link phase-biased by a superconducting ring pierced by an external magnetic flux, as schematically depicted in Figure \ref{fig:sketch}. The volume of the system is $V= A_N L_N + A_S L_S$ (where $A_{S/N}$ and $L_{S/N}$ represent the cross sectional area and the length of the superconducting/normal metal regions, respectively). In the following, we also denote with $\sigma_{S/N}$ the associated conductivities and with $\Delta_0$ the superconducting energy gap.

We assume that this hybrid system is thermally isolated. Its electronic degrees of freedom can be connected to two reservoirs residing at temperature $T_L$, $T_R$ via two ideal thermal valves $v_j$ with $j=L,R$, see Fig. \ref{fig:sketch}.
Ideal thermal valve means that they are assumed to be instantaneous, non dissipative and without losses.
We notice that a great effort in the research in the field of mesoscopic caloritronics\cite{karimi2016, strambini2014, ronzani2018, dutta2017,fornieri2017, sivre2018} is currently devoted to reach these conditions, developing novel schemes for an improved thermal isolation. In the following, we assume the valves closed (no heat exchange), except for Section \ref{sec:cycles} where the valves are exploited to realise thermodynamic cycles.

We consider a proximized SNS junction in the short junction regime, i.e. $L_N\ll \xi$, where $\xi$ is the superconducting coherence length. We also assume that the superconducting banks are long several coherence lengths ($L_S\gg \xi$), in order to impose BCS boundary conditions in the bulk\cite{hammer2007}. In the following we fix therefore $L_N=0.1\xi$ and $L_S=10\xi$. Numerical calculations are performed in the quasi 1-D approximation of Usadel equation\cite{hammer2007}, neglecting the edge-effects at the SN interfaces. This approximation is valid when $A_S\ll \xi^2$ or as long as the junction resistance is concentrated in the normal region\cite{golubov2004}. 

The proximity effect\cite{belzig1999, courtois1999} alters the DoS of both the weak-link and the superconductor. Qualitatively, in the normal metal appears an induced minigap $\bar{\Delta}(\ph)$ whose energy width can be tuned\cite{hammer2007,belzig1999,lesueur2008,giazotto2009} by the phase difference $\ph$. In a SNS junction, the minigap decreases monotonically by increasing the phase difference from $\varphi=0$ up to $\varphi=\pi$, where it completely closes $\bar{\Delta}(\varphi=\pi)=0$. Notice that $\bar{\Delta}(\ph)$ is symmetric for $\ph\rightarrow-\ph$ and $2\pi$-periodic, as usually happens for time-reversal symmetric proximized Josephson systems.

Since the DoS is phase-dependent, the quasiparticle entropy $S(T, \ph)$ of the junction acquires both a temperature $T$ and  phase drop $\ph$ dependence. 
The total entropy can be expressed as $S(T,\varphi)=\sum_{j=S,N}A_{j}\int~ dx_j {\cal S}(x_j,T,\varphi)$, where $x_j$ denotes the curvilinear coordinate along the superconducting ring ($x_S$) and normal region ($x_N$). $\cal S$ is the quasiparticle entropy density given by
\begin{equation}
\mathcal{S}(x,T,\varphi)  = -4\mathcal{N}_0 \int _{-\infty} ^{+\infty} N(x,\en,T,\varphi) f(\en,T)\ln f(\en,T) \dd \en \ ,
\label{eq:StatisticIntegral}
\end{equation}
where $f(\en,T)=1/(e^{\en/T}+1)$ is the Fermi distribution function (we set $\hbar =k_{{\rm B}}=1$), $N(x,\en,T,\ph)$ is the normalised local DoS that quantifies local variations due to the proximity effect, ${\cal N}_0$ is the normal DoS density at Fermi level.

The weak-link is in the diffusive regime, and we therefore use the quasiclassical Green function formalism to get the spatial and phase dependence of various quantities, such as DoS $N(x,\epsilon, T,\varphi)$ and entropy density ${\cal S}(x,T,\varphi)$. We thus numerically solve the associated Usadel equations as detailed in Ref.~\cite{virtanen2017, vischi2017} (see also Ref.~\cite{belzig1999,hammer2007} for technical details).
Since the entropy is given by the quasi-particle occupation of the available states, it increases from $\ph = 0$ (gapped state) to $\ph = \pi$ (gapless state), as depicted in Figure \ref{fig:two}(a). An analysis of the phase-modulation of the entropy for proximized SNS Josephson junctions can be found in Refs.\cite{rabani2008,rabani2009, virtanen2017}, where the relations between supercurrent, entropy and inverse proximity effect are taken into account. 

The entropy variation on phase and the supercurrent properties of the SNS junction are linked by a Maxwell thermodynamic relation\cite{virtanen2017}. Since both the entropy $S(T,\varphi)$ and the supercurrent $I(T,\ph)$ can be obtained as two different derivatives of the free energy of the system\cite{eilenberger1968,virtanen2017}, the following identity holds:
\begin{equation}
\frac{\partial S(T,\varphi)}{\partial \ph} = - \frac{1}{2 e }\frac{\partial I(T,\varphi)}{\partial T} \, \, ,
\label{eq:Maxwell}
\end{equation}
where $e$ denotes the electron charge. This identity is directly due to the equilibrium nature of the Josephson current, and it establishes an exact thermodynamic relation between the excited states (quasi-particles), responsible for the entropy, and the ground state properties characterized by the Cooper pairs, responsible for the supercurrent flow.

Equation (\ref{eq:Maxwell}) implies that the entropy can be written as
\begin{equation}
S(T, \ph)= S_0(T)+\delta S(T, \ph) \, \, ,
\label{eq:GeneralEntropy}
\end{equation}
being $S_0(T)$ the entropy at $\ph=0$ and
\begin{equation}
\delta S(T,\ph) = -\frac{1}{2e} \int _0 ^\ph \frac{\partial I(T, \ph')}{\partial T} \dd \ph' \, \, .
\label{eq:deltaS}
\end{equation}

From the above equations, asymptotic analytical limit can be easily obtained for a SNS diffusive junction. Indeed, for short weak link ($L_N/\xi \ll 1$) and $\sigma_S A_S/\sigma_N A_N \rightarrow \infty$, the current-phase-relation is given by the Kulik-Omel'yanchuk (KO) theory\cite{KO1975,golubov2004}. In the limit of the KO theory, the entropy at $\ph=0$ is $S_0(T)= V \mathcal{S}_{\rm BCS}(T)$ where $\mathcal{S}_{\rm BCS}$ is the homogeneous BCS entropy density, obtained by substituting in equation (\ref{eq:StatisticIntegral}) the normalised BCS DoS:
\begin{equation}
N(T, \en) = \left|
{\rm Re} \frac{\en}{\sqrt{\en^2-\Delta^2(T)}} \, \, 
\right| .
\label{eq:BCSDynesDoS}
\end{equation}
In this case, the entropy variation is\cite{heikkila2002,virtanen2017}
\begin{equation}
\delta S (T, \ph) = - \frac{2 {\cal N}_0 \Delta_0 V \alpha}{\pi}  \frac{\partial}{\partial T}  \int _{\DT \cos(\frac{\ph}{2})} ^{\DT} \tanh (\frac{\en}{2T})
\log[ \frac{\DT |\sin(\frac{\ph}{2})| + \sqrt{\en^2-\DT^2 \cos^2(\frac{\ph}{2})}}{\sqrt{\DT^2-\en^2}} ]  
\dd \en \, \, ,
\label{eq:KoModulation}
\end{equation}
where we have introduced the dimensionless parameter
\begin{equation}
\label{eq:alpha1}
\alpha= \frac{e R_0 I_c}{1.32 {\cal N}_0 \Delta_0^2 V}~=\pi^2 \frac{\xi}{L_N} \left(a\frac{L_S}{\xi}+\frac{L_N}{\xi} \right)^{-1} \, \, ,
\end{equation}
 with $R_0=\pi/e^2\approx \SI{12.9}{\kilo\ohm}$, $I_c=1.32 \pi \Delta_0/2 e R_N$ is the critical current at $T= 0$ in the KO theory\cite{golubov2004} and  $R_N$ is the normal-state resistance of the N region. In the above expression appears also the dimensionless parameter $a\equiv (A_S\sigma_S)/(A_N\sigma_N)$ which takes into account  geometric and electrical characteristic contributions of the S and N regions on proximity effect. Indeed, the parameter $a$ enters into the boundary conditions at the SN contact that requires the conservation of matrix current in a quasi-1D Usadel equations\cite{hammer2007,nazarov1994,virtanen2017}. Note that the last equality in equation (\ref{eq:alpha1}) has been obtained considering that the junction resistance is $R_N =  L_N/(\sigma_N A_N)$ and by using the relation\cite{virtanen2018} $ \xi^2 = \sigma_N/(2e^2{\cal N}_0 \Delta_0)$.
 
To appreciate the role of the $\alpha$  parameter it is convenient to introduce a quantity $p$ that estimates the relative variation of the entropy induced by the phase in comparison to the phase independent part, defined as
\begin{equation}
p(T) = \frac{S(T, \ph=\pi) - S(T,\ph=0)}{S(T,\ph=0)}=\frac{\delta S(T,\ph=\pi)}{S_0(T)} \, \, .
\end{equation}
The quantity $p(T)$ is reported in Figure \ref{fig:two}(b). The various curves correspond to different values of $a=(\sigma_S A_S)/(\sigma_N A_N)$, representing different geometrical and material configurations.
 
From the above equation one can see that, in the short junction limit, the relative entropy variation $p$ scales with $\alpha$ of equation (\ref{eq:alpha1}). Therefore, to increase the relative entropy variation, and thus making larger the effect due to iso-entropic process, one should increase $\alpha\propto I_C/V$, by increasing  the value of the critical current or by lowering the volume of the system $V$.

The adiabatic effects that we are going to study in the next sections depend on the entropy relative variation $p$ and hence on the parameter $\alpha$.
From equation (\ref{eq:alpha1}), obtained within KO theory, we could predict the expected scaling behaviour.  In particular, the magnitude of the entropy variation scales like $1/L_N$, since the critical supercurrent scales like $1/R_N$ for short junctions.
Instead, by increasing $L_S$ the total volume of the device increases without substantially affecting supercurrent, with the consequence that the relative entropy variation behaves like $1/L_S$. We note that 
this argument does not hold for $L_S \lesssim \xi$, where BCS rigid boundary conditions are not anymore valid and hence the supercurrent magnitude can depend on $L_S$. As already stated, here we neglect this situation by considering $L_S=10\xi$. It is important to notice, however, that the critical current-volume ratio cannot be increased at will: they are not independent, since for a small volume local self-consistent reduction\cite{riedel1996,cherkez2014}  of the pair potential $\Delta_0(x)$ will appear, decreasing the supercurrent of the weak-link.

It is important to stress that equations (\ref{eq:GeneralEntropy}) and (\ref{eq:deltaS}), and their link with the current-phase-relationship, are general and rely on basic thermodynamic consistency relation, i.e. Maxwell relation. Hence, they hold true independently of the nature of the weak-link (insulating barrier, metallic weak-link, ferromagnet layer, etc. ). We have performed calculations on a superconductor-insulator-superconductor (SIS) junction, using the Ambegaokar-Baratoff current-phase-relationship\cite{tinkham2004,golubov2004}, demonstrating that at the same supercurrent and volume the entropy modulation is suppressed compared to that of a SNS junction. This is due to the particular feature of the SNS junction that, thanks to the DoS of the N region, allows the phase-modulation of the correlations and transport properties over a volume of magnitude $\xi^3$, differently to a SIS junction that concerns a zero-length insulator layer. The above discussion shows the complete generality of the presented mechanism, which can be realized with different kind of junctions or with different external thermodynamic variables opening the road to further developments in the thermodynamic characterisation of novel hybrid systems.

\section{Iso-entropic process}
\label{sec:iso-entropic}
\begin{figure}
  \includegraphics[width=0.99\textwidth]{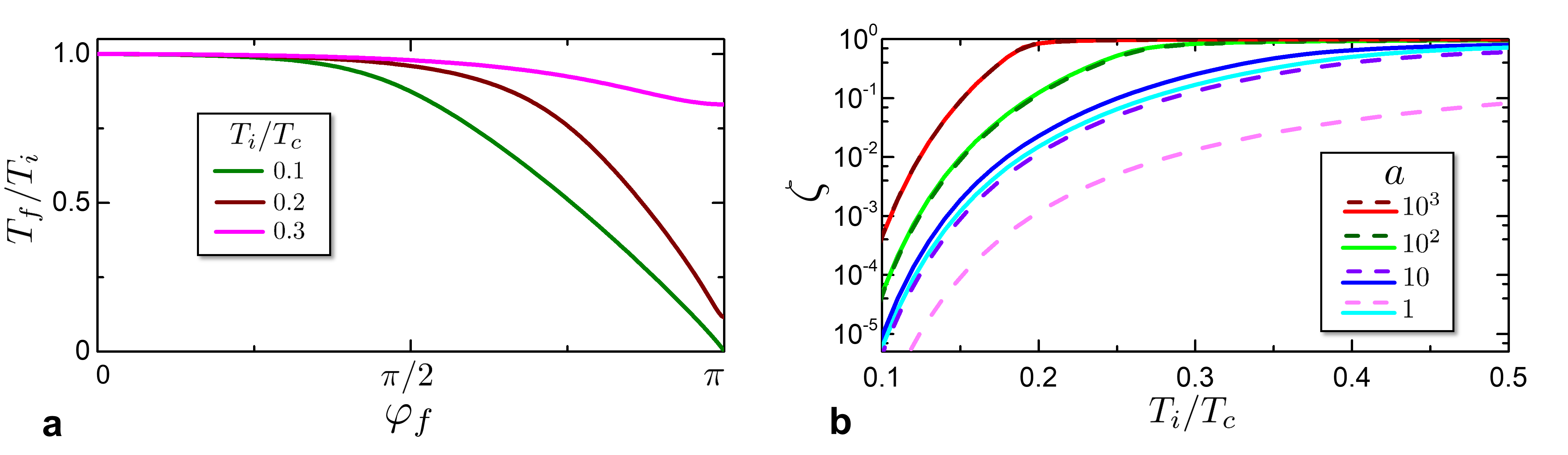}
  \caption{\textbf{ Temperature decrease in a iso-entropic process.}  (a) The relative temperature decrease $T_f/T_i$ as a function of phase $\ph_f$ of the junction, where $T_i$ is the initial temperature at $\ph_i=0$. Different curves refer to different initial temperature $T_i/T_c$ at a fixed $a=100$. (b) shows the ratio $\zeta(T_i)=T_f/T_i$ for an iso-entropic process $(T_i,\ph_i=0)\rightarrow (T_f,\ph_f=\pi)$, as a function of initial temperature $T_i$. Different curves refer to different values of $a$. Solid lines correspond to numerical solutions of Usadel equations; dashed lines correspond to analytic results obtained within the KO theory. The KO theory and the full numeric solution are in good agreement at $a=10^2,10^3$; deviations appear for lower values of $a$. }
\label{fig:TemperatureDecrease}
\end{figure}
In the previous section we have discussed the phase and temperature dependence of the thermodynamic entropy of a SNS junction. Exploiting these features, we now study the properties of an iso-entropic process in which the entropy remains constant, while externally varying the phase $\varphi$ of the weak-link. In order to retain the entropy constant, this process will result in a temperature variation, in particular in a electronic temperature decrease of the junction as we will demonstrate below. To implement such a process we assume that the system is thermally isolated, and does not exchange heat with the environment and phonons (see also Section \ref{sec:exp} for a detailed discussion of this issue in realistic experimental conditions). For this reason, in a single iso-entropic process the two thermal valves $v_L$ and $v_R$ sketched in Figure \ref{fig:sketch} are closed. Exploiting a physical analogy with classical thermodynamics, this iso-entropic process is similar to an adiabatic expansion/compression of an ideal gas. In both situations there is no heat exchange with an external reservoir and the number of available states is modified by the variation of a thermodynamical variable, typically an external parameter. In the former case tuning the phase $\varphi$ one can modify the value of the minigap (and consequently the DoS), while in the latter case varying the volume the available states will change.

Let us consider the system to be in a initial thermodynamic state $(T_i,\ph_i=0)$ with a entropy $S_i = S(T_i,\ph_i=0,)$. In an iso-entropic process, the entropy remains constant, thus, in a quasi-static process that brings the phase from $\ph_i=0$ to $\ph_f$, the final temperature is determined by the entropy equation $S(T_f,\ph_f) = S_i$. Here, a relation $T_f(\ph_f , S_i)$ between the temperature and the phase in the final state is implicitly established.
In particular, since the entropy increases from $\ph_i= 0 $ to $0<\ph_f\leq \pi$, the isolated system will decrease its temperature from its initial value $T_i$. This is shown in Figure \ref{fig:TemperatureDecrease}(a), where we plot the relative temperature decrease $T_f(\ph_f, T_i )/T_i$ as a function of $\ph_f$. Notice that greater temperature decrease can be achieved for lower initial temperature $T_i$ (see e.g. the curve corresponding to $T_i=0.1 T_c$ in Fig. \ref{fig:TemperatureDecrease}(a)).
Recalling the symmetry properties of the supercurrent\cite{likharev1979} and equation (\ref{eq:Maxwell}), it is possible to argue that $T_f(\ph_f , S_i)$ is  $2\pi$-periodic even function in $\ph_f$.

From now on we concentrate on a process that brings the junction from $\varphi_i=0$ to $\varphi_f=\pi$, as sketched with the black arrow in Fig. \ref{fig:two}(a), that maximize the temperature decrease. For these two values the supercurrent flowing through the junction is zero, allowing also to neglect the contribution to the system energy of the ring inductance (this point is further clarified in Section \ref{sec:cycles}). We define $\zeta (T) \equiv T_f/T$ for an iso-entropic process from $(\ph_i=0,T)$ to $(\ph_f=\pi,T_f)$. At low temperature ($T_i\lesssim 0.2T_c$), $\zeta(T_i)$ is characterised by an exponential decrease as a consequence of the energy mismatch between the distribution $f(\en)\log f(\en)$ and the proximized DoS with induced minigap $\overline{\Delta}$. Being the latter phase-dependent, when $\ph$ approaches $\pi$, the induced minigap closes and the energy window associated to  $f(\en)\log f(\en)$ becomes greater than the minigap. As a consequence, the phase-dependence of the entropy integral in equation (\ref{eq:StatisticIntegral}) is stronger at low temperatures. This property is reflected in the different behaviour of the entropy at $\ph=0$ and $\ph=\pi$ in Figure \ref{fig:two}(a).

In Figure \ref{fig:TemperatureDecrease}(b) we report $\zeta(T_i)$ as a function of $T_i$. Various curves refer to different geometries for different values of the dimensionless proximity parameter $a$. In this figure and the followings, the solid curves represent the numerical solution obtained by solving the Usadel equations\cite{virtanen2017,vischi2017} within the specified geometry (see Section \ref{sec:model}). The dashed lines are instead obtained within the KO theory, calculating of $S(T,\ph)$ by mean of equations (\ref{eq:StatisticIntegral}), (\ref{eq:BCSDynesDoS}), (\ref{eq:KoModulation}). In all cases, there is good agreement between the full numeric results and the KO theory for $a=10^2,10^3$, while deviations appear at lower values of $a$ where KO theory overestimates the temperature decrease.

\begin{figure}
  \includegraphics[width=0.99\textwidth]{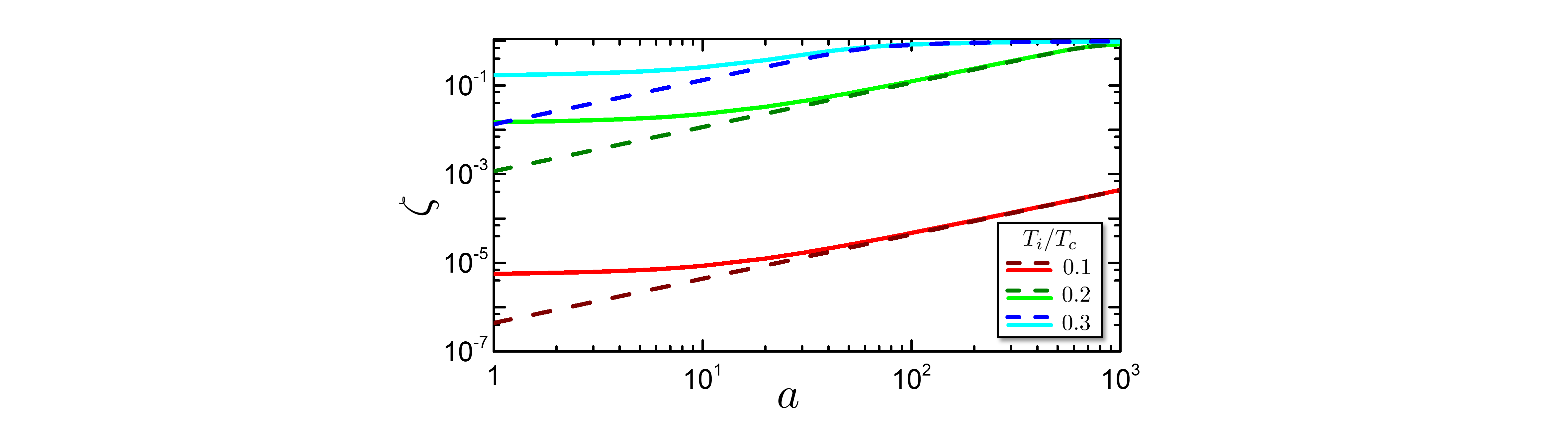}
  \caption{ \textbf{Ratio $\zeta=T_f/T_i$ as a function of $a$.} Different curves refer to different initial temperature $T_i/T_c$. Solid lines are calculated by numerically solving the Usadel equations. Dashed curves are analytical solutions within the KO theory, valid in the limit $a\gg 1$.
}
\label{fig:adiabatic}
\end{figure}
From Figure \ref{fig:TemperatureDecrease}(b) one can argue that $\zeta(T_i)$ grows with $a$. This is shown in detail in Figure \ref{fig:adiabatic}, where $\zeta(T_i)$ is plotted as a function of $a$ for three different values of initial temperature $T_i$. When $a \gg 1$, the current-phase-relationship of the junction tend to the KO asymptotic limit ($a\to \infty$) where the supercurrent magnitude is determined only by the junction geometry. 
In this limit, the entropy variation has the form of equation (\ref{eq:KoModulation}) with a scale determined by the critical supercurrent magnitude $I_c$. Furthermore increasing the volume $V$ of the system, $a$ increases and consequently the phase-independent contribution $S_0(T)$ of entropy increases. Then, the relative entropy variation $p$ decreases scaling with $\alpha$, spoiling the iso-entropic effects. In other words, the contribution $S_0$ that increases with $V$ acts as a heat-capacity that mitigate the iso-entropic temperature decrease. On the contrary, when $a\to 1$, the KO theory does not hold anymore. In particular, the current-phase-relationship is no more determined by the weak link characteristics only, but depends also on the geometrical parameters of the superconductor. This is due to the fact that the normal metal decreases the correlations in the superconductor banks, with the final result that a superconductor region nearby the SN interface behaves like a normal metal. As a consequence, the SNS junction behaves like a weak link with an effective length that is longer than the geometrical length\cite{likharev1979,riedel1996,Plehn1991} and with a reduced supercurrent magnitude that decreases with $a$. This effect contributes to decrease the iso-entropic effects, as shown by the fact that the numerical calculation (solid lines) in Figure \ref{fig:TemperatureDecrease} and \ref{fig:adiabatic} return a worse iso-entropic temperature decrease.

Analytical results in the KO theory for low temperature can be easily obtained. These results hold approximatively for $T\lesssim 0.1T_c$. Let us consider the entropy general form in equation (\ref{eq:GeneralEntropy}); at low temperature, the entropy at phase $\varphi=0$ (BCS form) has asymptotic behaviour\cite{abrikosov1965,degennes1966} 
\begin{equation}
S_0(T \to 0) \simeq \sqrt{2 \pi} \left(\frac{\Delta_0}{T}\right)^{1/2} e^{-\Delta_0/T} V \mathcal{N}_0 \Delta_0 \, \, .
\label{eq:BCSentropy}
\end{equation}

Instead, a linear-in-temperature behaviour can be found for $\delta S (T\to 0,\ph=\pi)$ of equation (\ref{eq:KoModulation}). At $\varphi=\pi$ we have
\begin{equation}
\delta S(T,\varphi=\pi)= \mathcal{N}_0 \Delta_0 V \frac{\alpha}{2 \pi}  \int_0^{\Delta_0}  \frac{\en}{T^2} \sech^2( \frac{\en}{2T})\ln\big(\frac{\Delta_0 + \en}{\Delta_0 - \en}\big)\dd \en~,
\end{equation}
and expanding the logarithm, we obtain
\begin{equation}
 \delta S(T\to0,\pi) 
 \simeq \mathcal{N}_0 T V \frac{\alpha}{\pi} \int_0^{\Delta_0/T}z^2\sech^2(z/2)\dd z \ ,
\end{equation}
 with $z=\en/T$. The low temperature limit $\Delta_0 / T \to \infty$ gives
\begin{equation}
 \delta S(T \to 0, \ph=\pi) \simeq  \frac{2 \pi }{3}\frac{e R_0 I_c}{1.32 \Delta_0} \frac{T}{\Delta_0} \, \, ,
 \label{eq:linear2}
\end{equation}
where one can see a linear-in-temperature behaviour reminescent of the normal metal nature at $\varphi=\pi$.
At low temperature the entropy variation $\delta S$ is proportional to the critical supercurrent $I_c$ of the junction. From equations (\ref{eq:linear2}) and (\ref{eq:BCSDynesDoS}) it is straightforward to verify  that the relative entropy variation $p(T)=\delta S(T,\pi)/S_0(T)$ scales like $\alpha$, as stressed in Section \ref{sec:model}.

In passing we notice that for a SIS junction, where the current-phase-relationship is given by the Ambegaokar-Baratoff formula \cite{tinkham2004,golubov2004}, the entropy variation at low temperature is
\begin{equation}
\delta S_{\rm AB}(T\to0,\ph=\pi) = \frac{e R_0 I_c}{2\pi} \frac{\Delta_0}{\cosh^2(\frac{\Delta_0}{2T})T^2} \, \, .
\end{equation}
Comparing this result with equation (\ref{eq:linear2}), we notice that the entropy variation in a SIS is exponentially suppressed with respect to that of a SNS junction, with a completely different temperature dependence. In order to enhance the effects of an iso-entropic transformation, therefore, it is convenient to consider a SNS junction instead of a SIS one.  

The temperature decrease $\zeta (T_i)$ can be obtained by solving numerically the transcendental equation $S_0(T_i) = S_0(T_f)+\delta S(T_f,\ph=\pi)$. However for $T\to 0$ the phase-independent contribution in equation (\ref{eq:BCSDynesDoS}) (with exponential behaviour in $T$) can be neglected respect to the linear term in equation (\ref{eq:linear2}). In this case, $S(T_f,\ph=\pi)\approx \delta S(T_f,\ph=\pi)$ and we obtain 
\begin{equation}
\zeta_{{\rm KO}} (T_i) \sim \frac{3}{\alpha\sqrt{2\pi}} \left( \frac{\Delta_0}{T_i}\right)^{3/2} e^{-\Delta_0/T_i} \, \, ,
\label{eq:gammaKO}
\end{equation}
where the subscript $\rm KO$ means that the equation is obtained within the KO theory. Writing $\alpha$ explicitly, it turns out that $\zeta_{\rm KO} (T_i ) \propto V/I_c$, confirming that the iso-entropic temperature decrease is enhanced by increasing the supercurrent and decreasing the volume. This is an important quantity because $\zeta (T_i) T_i$ represents the minimum achievable temperature in this iso-entropic transformation.

A comment on the heat capacity $C$ of the system is now in order.
This is a measurable quantity, encoding the temperature variation of the system after a given heat pulse, produced for example by Joule heating. Here, we can expect that the heat capacity is phase-dependent, being
\begin{equation}
C(T,\ph) = T \frac{\partial S(T,\ph)}{\partial T} \, \, .
\end{equation}
Hence, the two different temperature behaviours of entropy at $\ph=0$  and $\ph=\pi$ are reflected in the different behaviours of the heat capacity. Notice that here we consider the heat capacity at constant phase; this quantity can be different from the heat capacity at constant flowing supercurrent, in analogy with the different heat capacity of an ideal gas at constant volume or constant pressure. From equation (\ref{eq:BCSentropy}) we obtain that the heat capacity at $\ph=0$ is 
\begin{equation}
\label{eq:16}
C(T,\ph=0) = \sqrt{2 \pi} \left( \frac{\Delta_0}{T} \right)^{3/2} e^{-\Delta_0/T} V \mathcal{N}_0 \Delta_0 \, \, ,
\end{equation}
while the heat capacity at $\ph = \pi$ is $C(T,\ph=\pi)=C(T,\ph=0)+\delta C(T,\ph=\pi)$, where (see equation (\ref{eq:linear2}))
\begin{equation}
\delta C(T,\ph=\pi) = T \frac{\partial \delta S(T,\ph)}{\partial T} = \frac{2 \pi }{3}\frac{e R_0 I_c}{1.32 \Delta_0} \frac{T}{\Delta_0}\ .
\label{eq:deltaC}
\end{equation}
Similarly to the entropy, $C(T,\ph=\pi)\approx \delta C(T,\ph=\pi)$ at low temperature. Then, the heat capacity has a strongly different behaviour on $T/\Delta_0$ whether the phase is $\ph=0$ or $\ph=\pi$, respectively an exponential decrease or a linear behaviour with decreasing of temperature. Note that in the latter case one has the same linear-in-temperature behaviour expected for a normal metal, i.e. $C = 2\pi^2 \mathcal{N}_0 V T/3$. Notice that relative heat capacity variation $\delta C(T,\ph=\pi)/C(T,\ph=0)$ scales with $\alpha$. Hence, the value of $\alpha$ can be estimated by a heat capacity measurement at $\ph=0$ and $\ph=\pi$.

It is important to note that at low temperature the gapped nature of the superconducting leads exponentially suppress their heat capacity (see equation (\ref{eq:16})) producing a limited contribution with respect to the proximized region, even though they have larger in volume. This is one of the reasons behind the effectivity of the proposed iso-entropic transformation in affecting the electron temperature of the total system.

\section{Thermodynamic cycles}
\label{sec:cycles}
In this section we exploit the above results to implement thermodynamic cycles with a phase-biased Josephson weak-link. For sake of simplicity, here we limit the discussion to a particular cycle, although different thermodynamic cycles can be implemented. We thus combine two iso-entropic with two iso-phasic processes, in which $\ph$ is kept constant. In the following, we restrict to iso-phasic curves at $\ph=0$ and $\ph=\pi$. 

We investigate the Josephson Cycle properties and performances as an engine and a cooling system. These two different configurations depend on the orientation in which the processes are performed and which temperatures of the cycle are fixed by the reservoirs $T_L$ and $T_R$ as depicted in Figure \ref{fig:sketch}.

It is important to understand the form of the work and the heat associated to the weak-link during a  certain process.  We therefore fix the following sign convention: (I) the heat $Q$ absorbed by the system from the environment is positive, (II) the work $W$ released from the system to the environment is positive, (III) the sign of the supercurrent is positive when flowing in the direction of the phase gradient.

The work is done by an external magnetic field that induces a dissipationless current $I$ on the system. This work increases the energy of the system through two components. One is due to the reversible energy stored in the inductance that we can neglect since we treat the two states at $I=0$.
The second component represents the Josephson energy stored in the junction\cite{tinkham2004}. The work done on a junction from $\ph = 0$ to $\ph = \pi$ at constant temperature $T$ (i.e. in a iso-thermal process) is given by  $\frac{1}{2e} \int_0^\pi I(\ph,T) \dd \ph$, where $I(\ph,T)$ is the iso-thermal current-phase-relationship. In the case of an iso-entropic process, we must consider that the temperature is not constant. Considering the phase-dependence of the temperature  $T_f(\ph, T_i)$ (see Fig. \ref{fig:TemperatureDecrease}(a)), the work done by the system to the environment in an iso-entropic process $\ph=0\to\ph=\pi$ reads 
\begin{equation}
W_{i\to f} = - \frac{1}{2e}\int _0 ^\pi I_S(\ph,T_i) \dd \ph \, \, ,
\end{equation} 
where $I_S$ can be defined as
\begin{equation}
I_S (T_i, \ph)=I(\ph,T_f(\ph, T_i)) \, \, .
\end{equation}
In the following, we use the notation $W_{jl}$ to indicate the work done by the junction in a process from the thermodynamic state $j$ to the state $l$, where $j,l$ represent two states in Figure \ref{fig:engine}(a).

In the process $1\to2$ the work done by the system is negative, the environment must spend an amount of energy to charge the Josephson inductance of the junction, increasing the free energy of the latter\cite{tinkham2004,likharev1968,barone1982}. On the contrary, the system would release work when discharged from $\ph=\pi$ to $\ph=0$ (process $3\to4$). The work in a iso-phasic process is zero, being zero the phase variation, $W_{23}=W_{41}=0$.

At the same time, the heat absorbed by the junction in a process from the state $j$ to the state $l$ is
\begin{equation}
Q_{jl} = \int _j ^l  T \dd S \, \, ,
\label{eq:20}
\end{equation}
where $Q_{jl}=-Q_{lj}$.

In the entropy/temperature plane of Figure \ref{fig:engine}(a), the absolute value of heat $Q_{23}$ is  represented by the red area. The absolute value heat $Q_{41}$, instead is the sum of the red and blue area. In order to calculate the net work $W=W_{12}+W_{34}$ done in one cycle, we consider the conservation of energy which states that within a cycle $W$ is equal to the net heat absorbed by the system $Q = Q_{23}+Q_{41}$,  i.e. $W=Q$. 
Considering that $Q_{23}$ and $Q_{41}$ have opposite sign, the blue area in Figure \ref{fig:engine} (a) represents the net work $W$ per cycle.

\subsection{Josephson engine}
\begin{figure}[t]
  \includegraphics[width=0.99\textwidth]{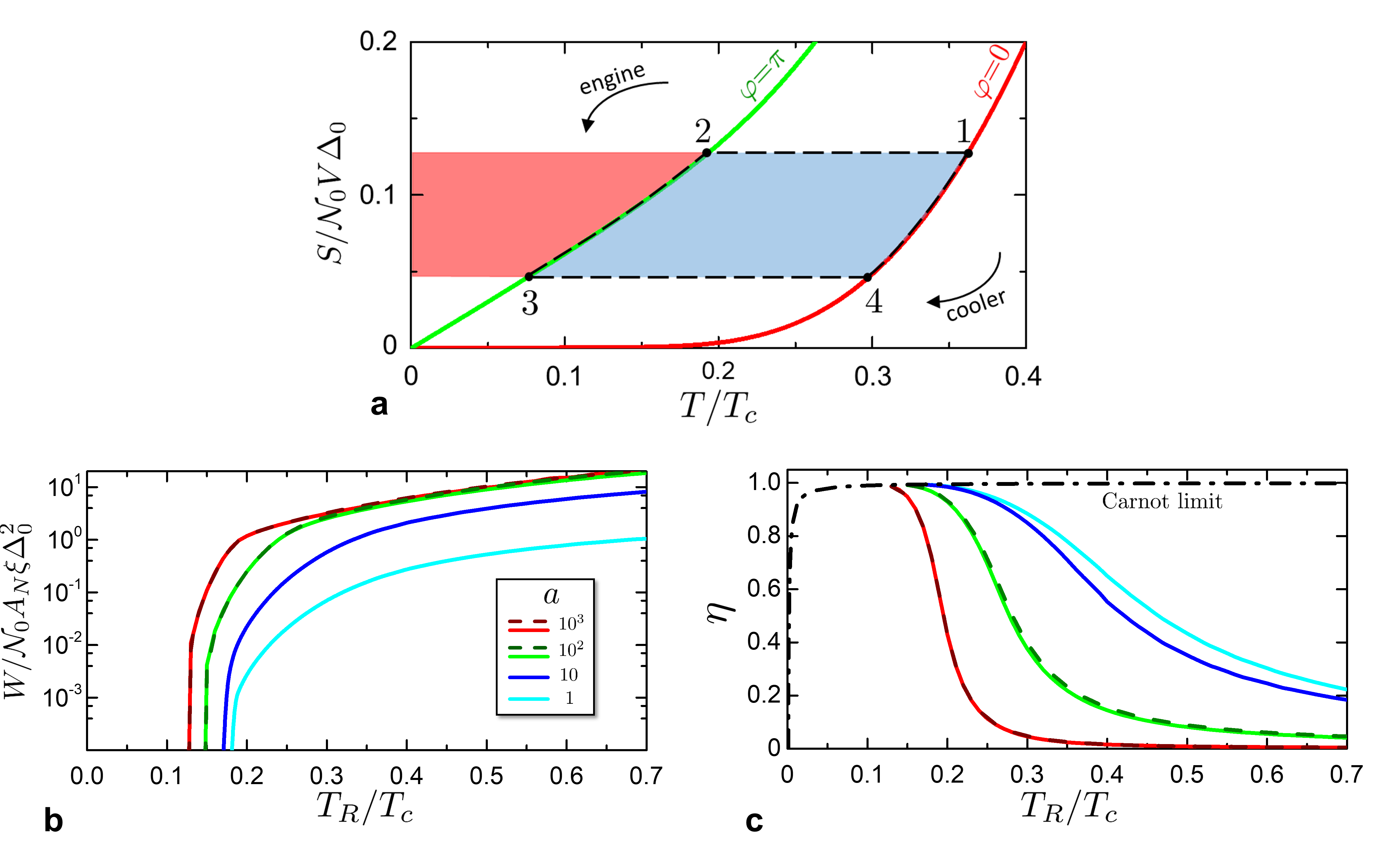}
  \caption{ \textbf{ SNS system as a thermodynamic machine.} (a) Scheme of a cycle in the entropy/temperature plane, consisting in the two iso-entropic processes $\mathbf{1 \text - 2}$ and $\mathbf{3 \text - 4}$ and the two iso-phasic processes $\mathbf{2 \text - 3}$ and $\mathbf{4 \text - 1}$. The red area represents the heat (released or absorbed) in the process $\mathbf{2 \text - 3}$. The light blue area is the work $W$ (done or received) in a complete cycle. The cycle can be used as engine or cooler machine if performed respectively counter-clockwise or clockwise. (b) Work $W$ of the engine as a function of the hot temperature $T_R$ at fixed $T_L=10^{-3} T_c$. Different curves refer to some values of $a$. Dashed curves are analytical solutions obtained within the KO theory. The curves are defined when $T_R \geq T_{\rm act}$. (c) Thermodynamic efficiency $\eta$ of the Josephson engine, for different $a$, as a function of $T_R/T_c$, with $T_L=10^{-3}T_c$. The black dot-dashed curve represents the Carnot limit $1- T_L/T_R$. The curves are defined for $T_R \geq T_{\rm act}$.
  }
  \label{fig:engine}
\end{figure}
An engine is a thermodynamic machine which can convert the temperature gradient between a hot and cold reservoirs into useful work. Referring to Figure \ref{fig:sketch}, we set the cold and hot reservoirs to be at temperature $T_L$ and $T_R$, respectively. The cold reservoir can be though as an environment ( e.g. large electric pad well-thermalised with the substrate at the base temperature of a cryostat) and the hot reservoir as a Joule heated pad with a continuous supplier of power. 
The Josephson engine consists of the following cycle, as sketched in Figure \ref{fig:engine}(a):
\begin{itemize}
\item {\bf Iso-entropic} $\mathbf{ 1 \rightarrow 2 .}$
The junction is initially at temperature of the hot reservoir $T_1= T_R$ and the valves $v_L, v_R$ are closed (thermally isolated junction). Then, an iso-entropic transformation brings the system from $(T_1=T_R,\ph_1=0)$ to $(T_2,\ph_2=\pi)$. The junction absorb a certain amount of work $|W_{12}|$ with no heat exchange.

\item {\bf Iso-phasic} $\mathbf{ 2 \rightarrow 3 .}$
The junction is put in contact with the cold reservoir by opening valve $ v_L$ and keeping the phase difference $\ph=\pi$. The iso-phasic transformation brings the system from $(T_2,\ph_2=\pi)$ to $(T_3=T_L,\ph_3=\pi)$. The junction releases heat $|Q_{23}|$ to the cold reservoir, without doing any work.

\item  {\bf Iso-entropic} $\mathbf{ 3 \rightarrow 4 .}$
The valve $v_L$ is closed and the junction is again thermally isolated. An iso-entropic transformation brings the system from
$(T_3=T_L,\ph_3=\pi)$ to $(T_4,\ph_4=0)$. The junction returns the work $W_{34}$ and no heat is exchanged.

\item {\bf Iso-phasic}  $\mathbf{ 4 \rightarrow 1 .}$
The junction is put in contact with the hot reservoir by opening valve $v_R$ and keeping the phase $\ph=0$. The iso-phasic transformation brings the system from $(T_4, \ph_4=0)$ to $(T_1=T_R,\ph_1=0)$. The junction absorbs heat $Q_{41}$ from the hot reservoir, without any work.
\end{itemize}
We fix the cold temperature to be $T_L=10^{-3}T_c$ (magnitude of \SI{}{\milli\kelvin}), and we study the engine performance as a function of $T_R$. Once fixed the cold reservoir  at $T_L$, it exists a minimum temperature threshold for the hot reservoir, below which the engine does not work. This threshold coincides with $T_4$, thus defined as activation temperature $T_{\rm act}=T_4$. This is implicitly defined by the equation $\zeta(T_{\rm act})T_{\rm act}=T_L$. When $T_R$ approaches $T_{\rm act}$, $T_2\to T_L$ and $T_R\to T_4$, with the consequence that the energy stored in the junction in the process $1\to2$ tends to the energy  returned in the process $3\to4$, and $W\to 0$. The activation temperature can be understood as follows: starting from the cycle in Figure \ref{fig:engine}(a), by decreasing $T_R$ till $T_{\rm act}$ the blue area collapses to a line, indicating that the work goes to zero.

Figure \ref{fig:engine}(b) shows the net work $W$ released by the junction as a function of $T_R$ for a fixed $T_L=10^{-3}T_c$. The curves are defined for $T_R \geq T_{\rm act}$. When $T_R \to  T_{\rm act}$ the $W$ curves go to zero. As one would expect, at fixed $a$, the net work is an increasing function of $T_R$, see that the blue area in Fig. \ref{fig:engine}(a) will increase if $T_1$ increases. A similar argument applies for the $a$ dependence of the net work, which tends to saturate for large values of $a$, while approaching the KO limit. On the other side the work decreases with decreasing $a$. The reason is that for small $a$, the effectivity of the iso-entropic transformation to change the device temperature is progressively affected  as a consequence of the reduction of the relative entropy variation $p$ (see discussion in Section \ref{sec:model}). The blue area of Fig. \ref{fig:engine} is reduced because transformation $2\to3$ and $4\to1$ become arbitrarily close, i.e. the neat work reduces.

Figure \ref{fig:engine} (c) reports the efficiency of the Josephson engine $\eta = W/Q_{41} = 1- |Q_{23}|/Q_{41}$. Also here, the various curves are defined for $T_R \geq T_{\rm act}$. The dash-dotted black curve is the Carnot efficiency limit given by $\eta_C = 1- T_L/T_R$.
For $T_R \rightarrow T_{\rm act}$ the efficiency tends to the Carnot limit. 
Considering Figure \ref{fig:engine} (a), the efficiency can be visualized as the ratio between the blue area and the total (red + blue) area. When the areas collapse to a line, the efficiency tends to $\eta = (T_{4}-T_3)/T_4=1-T_3/T_4$ that equals $\eta_C = 1- T_L/T_R$. This shows, for a Josephson engine, a common property shared with other engines that when the thermodynamical efficiency is maximal then the work produced tends to zero.

Analytical results can be obtained within the KO theory in the limit where temperatures are much smaller than $\Delta_0$. The value of $T_{\rm act}$ can be found by solving
\begin{equation}
T_L = \zeta_{KO}(T_{\rm act})T_{\rm act} =\frac{3\Delta_0}{\alpha\sqrt{2\pi}} \left( \frac{\Delta_0}{T_{\rm act}}\right)^{1/2} e^{-\Delta_0/T_{\rm act}} \, \, .
\end{equation}

By using equation (\ref{eq:20}), the heat released to the cold reservoir can be evaluated as 
\begin{equation}
\label{eq:q23}
Q_{23} = \int _{S_2} ^{S_3} T dS  = \frac{\pi }{3} \frac{eR_0 I_c}{1.32}\left[ 
\left( \frac{T_3}{\Delta_0} \right)^2  -  \left( \frac{T_2}{\Delta_0} \right)^2
\right]
,
\end{equation}

where we have used equation (\ref{eq:linear2}) which governs entropy at $\varphi=\pi$ for small $T$.
Substituting $T_2=\zeta(T_R)T_R$ and $T_3=T_L$,
\begin{equation}
Q_{23} = \frac{\pi }{3} \frac{eR_0 I_c}{1.32}\left[ 
\left( \frac{T_L}{\Delta_0} \right)^2 -  \left( \frac{\zeta(T_R) T_R}{\Delta_0} \right)^2 
\right] \, \, .
\label{eq:EngineReleasedHeat}
\end{equation}
For $T_L\to0$ using equation (\ref{eq:gammaKO}), we obtain that $|Q_{23}|\propto T_R^2 V^2/I_c$. As expected, the released heat increases with the temperature of the hot reservoir. The quadratic behaviour in $V$ is due to the fact that both the heat capacity and the temperature difference $T_2-T_L\approx \zeta(T_R)T_R$ increase with volume $V$. The behaviour as $1/I_c$ is determined by the product of the heat capacity prefactor at ($\varphi=\pi$), which grows with $I_c$ (see equation \ref{eq:deltaC}), with the temperature squared $T_2^2=(\zeta(T_R)T_R)^2\propto 1/I_c^{2}$.

The heat $Q_{41}=\int_{4}^{1} T dS$ absorbed by the system can be easily calculated using the Laisant theorem, which shows that $\int_{4}^{1} T dS+\int_{4}^{1} S\  dT=S_1(T_1) T_1-S(T_4)T_4$. 

Finally one finds $Q_{41}= S(T_R)T_R - S(T_4)T_4 - \int_{4}^{1}S dT$ where the last integral can be easily evaluated from the BCS free energy expression\cite{abrikosov1965,degennes1966}. Anyway for  $T_R\ll \Delta_0$ that integral may be safely neglected due to the exponential suppression of BCS entropy. Finally one finds

\begin{equation}
\label{eq:q41}
Q_{41}= \sqrt{2\pi} 
\left[
\left( \frac{T_R}{\Delta_0} \right)^{1/2} 
 e^{-{\Delta_0}/{T_R}}  - \left( \frac{T_4}{\Delta_0} \right)^{1/2} 
 e^{-{\Delta_0}/{T_4}} 
 \right] V \mathcal{N}_0 \Delta_0^2 \, \, .
\end{equation}
The work is given by $W=Q_{23}+Q_{41}$ and for $T_L\to 0$ we found
\begin{equation}
W=\sqrt{2\pi} \left(\frac{T_R}{\Delta_0}\right)^{1/2} e^{-\Delta_0/T_R} 
\left[ 
1- \frac{3}{2\alpha\sqrt{2\pi}} \left( \frac{\Delta_0}{T_R} \right)^{3/2} e^{-\Delta_0/T_R} 
\right]
V \mathcal{N}_0 \Delta_0^2 \, \, .
\label{eq:work}
\end{equation}
This expression is valid because $S(T,\ph=\pi)\approx \delta S(T,\ph=\pi)$ in the low temperature limit. Since the second term in the square brackets is proportional to $1/\alpha$, the work slightly increases with supercurrent as $\sim 1-\kappa V/I_c$ where $\kappa$ is a factor exponentially suppressed at low temperatures.
The work follows roughly the temperature scaling $T_R^{1/2} e^{-\Delta_0 /T_R}$. 

Finally, the efficiency is given by
\begin{equation}
\eta=1-\frac{3}{2\alpha{\sqrt{2\pi}}} \left( \frac{\Delta_0}{T_R} \right)^{3/2} e^{-\Delta_0/T_R} \, \, ,
\end{equation}
and it decreases with temperature and  $\alpha^{-1}\propto V/I_c$. From this rough estimation  we expect that indeed it increases while lowering $a$ for a fixed $T_R$, and this is indeed found in the full numerical solution plotted with solid lines in Figure \ref{fig:engine}(c). 

\subsection{Josephson cooler}
\begin{figure}
  \includegraphics[width=0.99\textwidth]{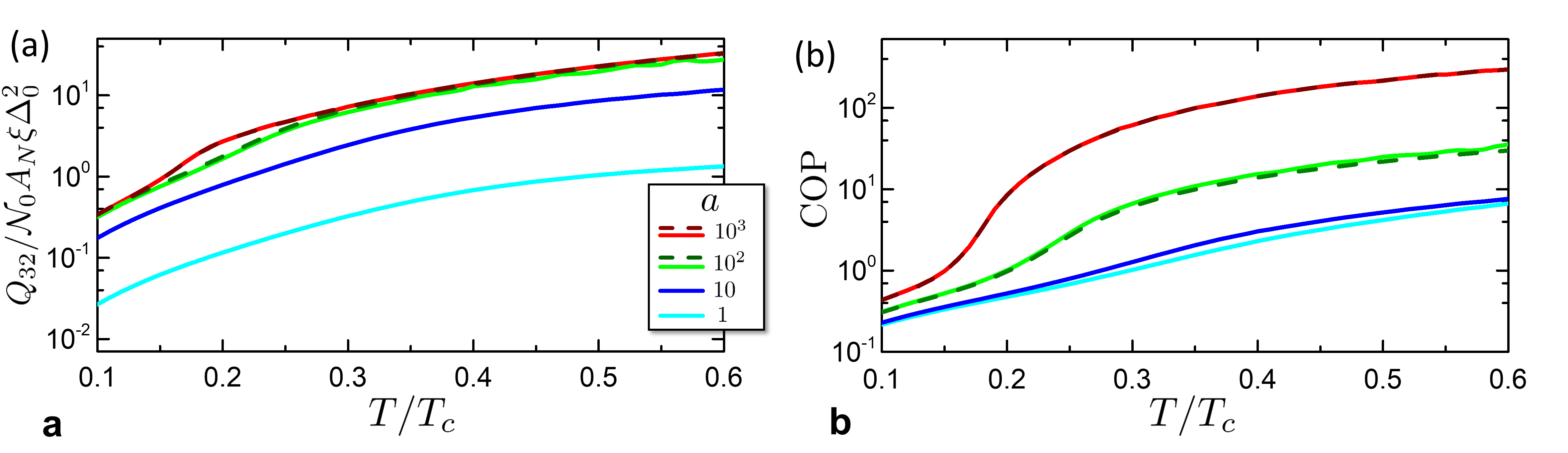}
  \caption{ \textbf{ Josephson cooler characteristics.} (a) Cooling Power per cycle $Q_{32}$ of the refrigerator at cooling fin temperature equal to the environment temperature $T_L=T_R$, as a function of $T_R/T_c$ and for different values of $a$. Solid curves are obtained by numerical solution while dashed curves represent analytical results obtained within the KO theory.
(b) Coefficient of performance of the cooler at $T_R=T_L$ for various $a$. 
  }
  \label{fig:cooler}
\end{figure}
We now discuss the thermodynamic cycle in the opposite configuration, i.e. acting as a cooler.
Indeed, cooling can be obtained by reversing the cycle described in the previous section and by considering that the role of the reservoir is different in this situation. 

We consider the junction to be connected to a reservoir at temperature $T_R$ via a thermal valve $v_R$, and to an external system --to be cooled-- initially at temperature $T_L$, named cooling-fin in the following. The latter system has to be thermally isolated from other spurious heat sources. This may be realised in nanoscaled suspended systems, such as membranes\cite{clark2005}, circuits\cite{koppinen2009} and low dimensional electronic systems.

The Josephson cooling cycle consists in the following four processes
\begin{enumerate}
\item  $\mathbf{ 4 \rightarrow 3 .}$
The junction is in equilibrium at environment temperature $T_4 = T_R$ and the thermal valves $ v_L$, $ v_R$ are closed.
An iso-entropic transformation brings the system from $(T_4=T_R,\ph_4=0)$ to $(T_3,\ph_3=\pi)$ doing work over the junction.
\item  $\mathbf{ 3 \rightarrow 2 .}$
The thermal valve $v_L$ is open, making the junction in contact with the cooling fin, reaching $T_L$. In this process, the junction absorb heat $Q_{32}$ from the cooling fin.
\item  $\mathbf{ 2 \rightarrow 1 .}$
The thermal valve $ v_L$ is closed. An iso-entropic transformation brings the system from
$(T_2=T_L,\ph_2=\pi)$ to $(T_1,\ph_1=0)$. In this process, work is released from the junction to the external circuit and no heat is exchanged. The temperature $T_1$ represents the higher temperature of the cycle and is the analogue of the temperature in the heat exchanger in refrigerators.
\item  $\mathbf{ 1 \rightarrow 4 .}$
The thermal valve $v_R$ is open, making the junction in contact with the environment. The junction temperature is lowered from $T_1$ to $T_R$, releasing heat to the environment.
\end{enumerate}

The temperature $T_3$ is again determined by the iso-entropic cooling from the temperature of the environment $T_4=T_R$. The temperature $T_3$ is the minimum possible temperature of the cooling-fin, i.e. the minimum achievable temperature $T_{\rm MAT}(T_R)$ once given $T_R$ as the one of the hot reservoir.
The $T_{\rm MAT}$ is given by an iso-entropic transformation starting from the state $(T_R,\varphi=0)$ to $(T_{{\rm MAT}},\varphi=\pi)$. We thus recall analytic result for $T_{\rm MAT}$ that holds at $T_R \ll \Delta_0$ in the KO theory:
\begin{equation}
T_{\rm MAT} = \zeta_{{\rm KO}} (T_R) T_R = \frac{3\Delta_0}{\alpha\sqrt{2\pi}} \left( \frac{\Delta_0}{T_R}\right)^{1/2} e^{-\Delta_0/T_R} \, \, .
\end{equation}
If $T_L=T_{\rm MAT}$, no heat can be absorbed by the junction from the cooling-fin. 

To characterize the performance of the Josephson cooler, we discuss the cooling power per cycle $Q_{32}$ and the coefficient of performance $\rm COP$ of the SNS junction when the cooling-fin temperature equals the temperature of the environment, i.e. $T_L=T_R=T$. 
In this state the cooling power is maximum and decreases to zero when the cooling-fin temperature approaches  the $T_{\rm MAT}$.
The cooling power per cycle $Q_{32}$ represents the amount of heat removed from the cooling-fin and is plotted in Figure \ref{fig:cooler} (a). Indeed the iso-entropic cooling process reduces the device temperature lower than the cooling fin and by thermalization the heat is absorbed by the cooler from the cooling-fin. The cooling power then depends on the heat capacity of the system during the thermalisation process ($3\to2$ in Fig. \ref{fig:engine}).
When the system passes from the state $4$  to the state $2$ with the same temperature $T$ with the phase difference tuned from $0$ to $\pi$, a certain amount of heat must be absorbed by the system, since the heat capacity is increased by $\delta C = T \partial _T \delta S$ (see Section \ref{sec:iso-entropic}). This quantity scales with the supercurrent magnitude and for $a\to \infty$ converges to a function defined by the KO current-phase relationship. 
This quantity can be estimated for $T\ll\Delta_0$ as
\begin{equation}
Q_{32} = \frac{\pi}{3} \frac{e R_0 I_c}{1.32} (1- \zeta_{\rm KO}(T)^2)\frac{T^2}{\Delta_0^2} \, \, .
\end{equation}
The dependence on $a$ is encoded in $\zeta_{{\rm KO}}(T)$. Neglecting the second term in the square brackets (since $\zeta(T)\to 0$ for $T\to 0$), the cooling power per cycle is 
\begin{equation}
\label{eq:q32}
Q_{32} = \frac{\pi}{3} \frac{e R_0 I_c}{1.32} 
\frac{T^2}{\Delta_0^2} \, \, .
\end{equation}
Intriguingly this limit does not depend on $a$ but only on the junction critical supercurrent.
Looking at Figure \ref{fig:cooler} (a), one see from the full numerical solution (solid lines) that $Q_{32}$ increases with $a$ and they tend to saturate to a particular limit for large values of $a$.
This trend is a consequence of the limited value of the critical current in the KO limit ($a\to \infty$). Indeed the cooling power increases with $a$ due to the fact that the junction effective lengths decrease when approaching the KO limit with maximal critical current.\cite{likharev1979,riedel1996,Plehn1991}  

The coefficient of performance (COP) is defined as the ratio of the pumped heat per cycle with the work spent per cycle, i.e. ${\rm COP}=|Q_{32}|/W$. We note that in analogy with the efficiency $\eta$ of an engine, also for this quantity there is a maximal limit ${\rm COP}_C$ that depend on the temperature of the two reservoirs, given by ${\rm COP}_C = T_L/(T_R-T_L)$.
In the specific case we are considering here with $T_R=T_L$ the ${\rm COP}_C$ diverges, therefore any  finite value of COP may be considered as a sort of inefficiency of the cooler with respect to an ideal one.

In Figure \ref{fig:cooler} (b) we show the COP as a function of the working temperature. One immediately see that the maximal performance of the cooler is obtained for $a\to\infty$. This corresponds to the maximal value of the absorbed work. At the same time the work absorbed per cycle decreases with $a\to \infty$. The reason is the following: increasing $a$, $T_1\to T_4$ and $T_3\to T_2$. Hence, the iso-entropic electric work $-W_{43}$ done on the system  tend to $W_{21}$, with the consequence that the net work per cycle $W=W_{43}+W_{21}$ tend to zero. For this reason, the COP increases with $a$, as reported in Fig. \ref{fig:cooler} (b).

It is possible to evaluate an asymptotic expression of the COP as a function of the temperature $T_1$ by solving the implicit equation $T=\zeta_{{\rm KO}}(T_1)T_1$. In the following calculation, we took for $Q_{32}$ the estimate of equation (\ref{eq:q32}) and for $Q_{14}$ we apply the same approach used to derive equation (\ref{eq:q41}) but neglecting the contribution of $T$ since in the limit $T\to 0$ one has $T\ll T_1$. So the heat exchange of the system with the hot reservoir  is
\begin{equation}
Q_{14}= -\sqrt{2\pi} 
\left( \frac{T_1}{\Delta_0} \right)^{1/2} 
 e^{-{\Delta_0}/{T_1}}  V \mathcal{N}_0 \Delta_0^2 \, \, ,
\end{equation}
 where the minus sign indicates that the heat is released by the system to the reservoir.

This rough estimates allows to inspect the scaling properties of the cooling cycle COP on $T_1$ and $\alpha$. The COP can be written as
\begin{equation}
{\rm COP} = -\frac{Q_{32}}{Q_{32}+Q_{14}}=\frac{1}{\left |\frac{Q_{14}}{Q_{23}}\right|-1} \ ,
\end{equation}
where
\begin{equation}
\left |\frac{Q_{14}}{Q_{23}}\right| = \alpha\frac{(2\pi)^{3/2}}{3\pi} \left(\frac{T_1}{\Delta_0}\right)^{{3/2}} e^{\Delta_0/T_1}  \, \, .
\end{equation}
Finally we have 
\begin{equation}
{\rm COP} =\frac{1}{2\sqrt{2\pi}\alpha (T_1/\Delta_0)^{{3/2}} e^{\Delta_0/T_1}/3-1} \, \, .
\end{equation}
At low temperatures the COP behaves as $\alpha^{-1}(T_1/\Delta_0)^{{-3/2}}e^{-\Delta_0/T_1}$ so, it decreases with temperatures decreasing. At the same time we expect to see that it grows with $a$ as suggested by the dependence of $\alpha$, i.e. equation (\ref{eq:alpha1}). All these behaviours are indeed seen in Figure \ref{fig:cooler}(b) where the solid lines represent the full numerical results and the dashed lines the results obtained within KO theory. Interestingly, increasing the supercurrent the COP decreases as $1/I_c$. The reason is that, increasing the supercurrent, the work scales as a power greater than the cooling power per cycle. This is due to the enhanced adiabatic decreasing of temperature. Therefore, improving the cooling power will result in a minor coefficient of performance, a common feature shared by refrigerators.

\section{Possible experimental implementations}
\label{sec:exp}
Here, we clarify which are the main physical requirements that need to be satisfied in order to realise the proposed thermal machines and we give some estimations on the expected performances.

The first important issue is to show the possibility to realise an iso-entropic transformation, i.e. to thermally isolate the electron system of the SNS junction from the thermal bath for the time necessary to perform the transformation. In these metallic system, electrons thermally relax mainly by electron-electron and  electron-phonon interactions, with respectively the two characteristic time scales $\tau_{e-e}$ and $\tau_{e-ph}$. An efficient iso-entropic process should be faster than the electron-phonon relaxation time and slower than the electron-electron relaxation time, keeping the electron system in thermal
quasi-equilibrium. This condition can be achieved in typical superconductors, where the two time scales are well-separated at low temperature\cite{kaplan1976}, with $\tau_{e-e}$ order of magnitude smaller\cite{kaplan1976} than the $\tau_{e-ph}$. For example, below $0.1 T_c$,  $\tau_{e-ph} $  can range from $\sim 100$ns of niobium-based superconductors\cite{kardakova2015,guarcello2017} to $\sim 10^1 \div 10^2\SI{}{\micro\second}$ of aluminium or tantalum thin films\cite{barends2008,timofeev2009}, while
$\tau_{e-e}$ is about $1\div 10$ns\cite{zgirski2017}. Then iso-entropic processes are possible with phase-switching rates in the window $10\div10^3$\SI{}{\mega\hertz}.

In view of possible experimental realizations, we can give some estimates on the expected performance of the proposed thermal machines, based on state-of-the-art materials and experimental parameters.
We thus consider experimentally reasonable values of $I_c=\SI{1}{\milli\ampere}$ and cycling frequency $\nu=\SI{100}{\mega\hertz}$.
At low temperature ($T/T_c\approx 0.2$) one would estimate a cooling power about \SI{2}{\pico\watt}. A similar order of magnitude is predicted for the released power, when the thermal machine is operated as a engine.

A possible implementation for thermal valves can be realized with quantum point contacts realized on top of two dimensional electron gas,  offering the high tunability and the degree of thermal isolation required to test our predictions\cite{1,2,4,5}. In this case good thermal contact between the two dimensional electron gas and the superconductor ring can be achieved using InAs-based quantum well and Nb or Al as superconductors\cite{1,2,3,4,5}  providing very high transparencies of the interfaces. A rough estimate of the expected temperature reduction for a realistic setup where the cooling-fin is done by a two-dimensional electron gas can be done. In III-V semiconductor crystals at sub-Kelvin temperatures the electron-phonon piezoelectric coupling is the dominant process for the heat exchange between electrons and the environment\cite{Price82,Gasparinetti11}. Then, for a realistic setup in which the  cooling fin is made by a quantum well, the heat transferred by the hot phonons to the electrons can be written as $\dot{Q}_{e-ph}(T,T_{ph})=\Sigma \mathcal{A}(T_{ph}^5-T^5)$ with
$\mathcal{A}$ the area of the quantum well and $T$ and $T_{ph}$ are the electron and phonon temperatures, respectively.
The coupling constant\cite{Gasparinetti11} has the typical values of
$\Sigma\approx 30$~\SI{}{\femto\watt} \SI{}{\micro\meter}$^{-2}$ \SI{}{\kelvin}$^{-5}$. The equilibrium electronic temperature $T^*$ of the cooling-fin can be easily obtained by solving the heat balance equation $\dot{Q}_{e-ph}(T^*,T_{ph})=\dot{Q}_{\rm cool}(T^*)$ where the cooling power of the system is $\dot{Q}_{\rm cool}(T)= Q_{32}(T) \nu$ with $Q_{32}$ taken from equation (\ref{eq:q32}) and $\nu$ is the operating frequency of the cooling cycle.
Using the previous values and assuming for the cooling-fin the area $\mathcal{A}\approx 100$~\SI{}{\micro\meter}$^{2}$ with a phonon temperature of $T_{ph}\approx 100$~\SI{}{\milli\kelvin} one finds for the equilibrium electron temperature $T\approx 1$~\SI{}{\milli\kelvin}. If $T\ll T_{ph}$  the solution of the balance equation may be approximated with $T\approx\sqrt{T_{ph}^5/\beta}$ with
$\beta=(\pi/3) (e R_0 I_c/1.32)(\nu/\Delta_0^2 \Sigma \mathcal{A})$.

This discussion shows the potential of this cooling cycle and demonstrates that the proposed thermodynamic cycles could operate efficiently in the sub-Kelvin regimes playing an important role for many different quantum technology platforms. 

\section{Summary and conclusions}
\label{sec:conclusion}
In this work, we have considered the thermodynamic properties of a proximized SNS Josephson junction in the diffusive regime. We have shown that the phase- and temperature-dependent entropy can be exploited to achieve significant temperature decrease of the electronic degrees of freedom of the system. In particular, one can implement iso-entropic processes, by externally tuning the phase drop of the weak-link, getting temperature variations consistent with thermodynamic constraints. Elaborating on this concept, we have demonstrated the possibility to build thermodynamic cycles based on the combination of iso-entropic and iso-phasic processes. By coupling the SNS junction to two thermal baths via two thermal valves, we have shown that it is possible to engineer a Josephson engine and cooler by coherently driving the phase across the weak-link. We have studied in detail these thermal machines, investigating their performances such as the efficiency or the cooling power as a function of different geometrical and electrical parameters. Full numerical calculations have been supported by asymptotic calculations valid in the short junction regime within the KO theory, discussing several limiting behaviours. We have also proposed a possible experimental setup to implement the discussed device  as powerful cooler at  sub-Kelvin regimes.

\section*{Acknowledgments}
We thank G. Marchegiani for useful discussions. F.V. , P.V. , A.B., E.S. and F.G. acknowledge
funding by the European Research Council under the European Union's
Seventh Framework Program (FP7/2007-2013)/ERC Grant agreement
No. 615187-COMANCHE. 
M.C. acknowledges support from the Quant-Era project "SuperTop". A.B. and M.C. acknowledge the support of  CNR-CONICET cooperation programme ``Energy conversion in quantum, nanoscale, hybrid devices''. A.B. acknowledges the Royal Society through the International Exchange between UK and Italy (grant IES R3 170054).  
F. G. acknowledges funding by Tuscany Region under the FARFAS 2014 project SCIADRO.

\section*{Author contributions statement}

F.V., M.C., P.V., E.S., A.B.,F.G. conceived the ideas and designed the study. F.V. developed the algorithm and carried out the simulations. F.V. and M.C. analyzed numerical results and evaluated asymptotic expansions. F.V., M.C., P.V., E.S., A.B., and F.G discussed the results and wrote the paper.

\section*{Additional information}

The authors declare no competing interests.

\bibliographystyle{naturemag}

\end{document}